\documentclass[letters,usenatbib]{mnras}

\usepackage{newtxtext,newtxmath}

\usepackage[T1]{fontenc}
\usepackage{ae,aecompl}

\usepackage{graphicx}	% Including figure files
\usepackage{amsmath}	% Advanced maths commands
\usepackage{amssymb}	% Extra maths symbols
\usepackage{booktabs}	% Add notes to tables
\usepackage{ragged2e}	% Commands for text justification

\title[Episodic outflow in the water fountain IRAS~18113--2503]{Rapidly-evolving episodic outflow in IRAS~18113--2503: clues to the ejection mechanism of the fastest water fountain}

\author[Orosz et al.]{G.~Orosz,$^{1,2}$\thanks{E-mail: gabor@orosz.space}
J.\,F.~G\'omez,$^{3}$\thanks{On sabbatical leave at Joint ALMA Observatory, Santiago, Chile}
H.~Imai,$^{4,5}$
D.~Tafoya,$^{6}$
J.\,M.~Torrelles,$^{7,8,5}$
R.\,A.~Burns,$^{9}$
\newauthor
P.~Frau,$^{10}$
M.\,A.~Guerrero,$^{3}$
L.\,F.~Miranda,$^{3}$
M.\,A.~Perez-Torres,$^{3}$
G.~Ramos-Larios,$^{11}$
\newauthor
J.\,R.~Rizzo,$^{12}$
O.~Su\'arez,$^{13}$
and L.~Uscanga$^{14}$
\\
\scriptsize{$^{1}$Xinjiang Astronomical Observatory, Chinese Academy of Sciences, 150 Science 1-Street, Urumqi, Xinjiang 830011, China}\\
\scriptsize{$^{2}$Konkoly Observatory, MTA Research Centre for Astronomy and Earth Sciences, Konkoly Thege Mikl\'os \'ut 15-17, Budapest 1121, Hungary}\\
\scriptsize{$^{3}$Instituto de Astrof\'{\i}sica de Andaluc\'{\i}a, CSIC, Glorieta de la Astronom\'{\i}a s/n, E-18008 Granada, Spain}\\
\scriptsize{$^{4}$Center for General Education, Institute for Comprehensive Education, Kagoshima University, 1--21--30 Korimoto, Kagoshima 890--0065, Japan}\\
\scriptsize{$^{5}$Department of Physics and Astronomy, Graduate School of Science and Engineering, Kagoshima University, 1--21--35 Korimoto, Kagoshima 890--0065, Japan}\\
\scriptsize{$^{6}$Chile Observatory, National Astronomical Observatory of Japan, National Institutes of Natural Science, 2--21--1 Osawa, Mitaka, Tokyo, 181--8588, Japan}\\
\scriptsize{$^{7}$Institut de Ci\`encies de l'Espai (ICE, CSIC), Can Magrans s/n, E-08193, Cerdanyola del Vall\`es, Catalonia}\\
\scriptsize{$^{8}$Institut d'Estudis Espacials de Catalunya (IEEC), E-08034, Barcelona, Catalonia}\\
\scriptsize{$^{9}$Joint Institute for VLBI ERIC (JIVE), Postbus 2, NL-7990 AA Dwingeloo, the Netherlands}\\
\scriptsize{$^{10}$Private researcher}\\
\scriptsize{$^{11}$Instituto de Astronom\'{\i}a y Meteorolog\'{\i}a, CUCEI, Universidad de Guadalajara, Av. Vallarta No. 2602, Col. Arcos Vallarta, 44130 Guadalajara, Jalisco, Mexico}\\
\scriptsize{$^{12}$Centro de Astrobiolog\'{\i}a (INTA-CSIC), Ctra.~M-108, km.~4, E-28850 Torrej\'on de Ardoz, Madrid, Spain}\\
\scriptsize{$^{13}$Universit\'{e} C\^{o}te d'Azur, Observatoire de la C\^{o}te d'Azur, CNRS, Laboratoire Lagrange, Bd de l'Observatoire, CS 34229, 06304 Nice cedex 4, France}\\
\scriptsize{$^{14}$Departamento de Astronom\'ia, Universidad de Guanajuato, A.P. 144, 36000 Guanajuato, Gto., Mexico}
}

\date{Accepted 2018 September 17. Received 2018 September 17; in original form 2018 June 15.}

\pubyear{2018}

\begin{document}
\label{firstpage}
\pagerange{\pageref{firstpage}--\pageref{lastpage}}
\maketitle

% Abstract of the paper
\begin{abstract}
Water fountains are evolved stars showing early stages of collimated mass loss during transition from the asymptotic giant branch, providing valuable insight into the formation of asymmetric planetary nebulae. We report the results of multi-epoch VLBI observations, which determine the spatial and three-dimensional kinematic structure of H$_2$O masers associated with the water fountain IRAS~18113$-$2503. The masers trace three pairs of high-velocity ($\sim$150--300~km~s$^{-1}$) bipolar bow shocks on a scale of 0$\farcs$18 ($\sim$2000~au). The expansion velocities of the bow shocks exhibit an exponential decrease as a function of distance from the central star, which can be explained by an episodic, jet-driven outflow decelerating due to drag forces in a circumstellar envelope. Using our model, we estimate an initial ejection velocity $\sim$840~km~s$^{-1}$, a period for the ejections $\sim$10~yr, with the youngest being $\sim$12~yr old, and an average envelope density within the H$_2$O maser region $n_{\text{H}_2}$$\approx$$10^{6}$~cm$^{-3}$. We hypothesize that IRAS~18113$-$2503 hosts a binary central star with a separation of $\sim$10~au, revealing novel clues about the launching mechanisms of high-velocity collimated outflows in water fountains.
\end{abstract}

% Select between one and six entries from the list of approved keywords.
% Don't make up new ones.

%%%%%%%%%%%%%%%%%%%%%%%%%%%%%%%%%%%%%%%%%%%%%%%%%%

%%%%%%%%%%%%%%%%% BODY OF PAPER %%%%%%%%%%%%%%%%%%

\section{Introduction}
\label{sec:introduction}

A long-standing puzzle in the late stellar evolution of low- to intermediate-mass stars ($\sim$1--8~M$_\odot$) is the shaping mechanism of asymmetric planetary nebulae (PNe). It involves a sudden change in the stellar mass-loss mode from spherical to bipolar/multipolar that occurs in a few hundreds of years between the late asymptotic giant branch (AGB) and post-AGB phases, before forming PNe \citep[e.g.,][]{balickfrank2002}. This transition is characterized by the launching of high-velocity collimated outflows, which then interact with and carve out aspherical cavities in the slow winds of the circumstellar envelope (CSE) ejected during the AGB phase \citep{sahaitrauger1998,sahai2005,sahaipatel2015}. These structures are subsequently ionized by stellar radiation to form the signature appearance of young PNe. Unraveling the origin and launching mechanisms of high-velocity collimated outflows in evolved stars requires studying them at an early stage after the launching events \citep[e.g.,][]{bujarrabal2001,vlemmings2014}.

A group of sources in such an early stage are water fountain stars (WFs), evolved objects in transition between the late AGB and early PNe that show fast 22~GHz H$_2$O maser emission tracing post-shock gas at the interfaces between jet-driven outflows and their surrounding CSEs \citep{imai2007,desmurs2012,gomez2017}. Outflow speeds are usually a few hundred km~s$^{-1}$, about an order of magnitude faster than the slowly expanding CSE formed in the AGB phase. Collisionally excited H$_2$O masers make it possible, by using very long baseline interferometry (VLBI), to characterize the spatio-kinematic properties of the outflows. VLBI measurements revealed that H$_2$O masers trace very recent outflow activity in WFs, usually only a few decades old, which can be collimated by large-scale magnetic fields \citep{imai2002,vlemmings2006}.

IRAS 18113$-$2503 (hereafter I18113) was identified as a post-AGB WF by \citet{gomez2011}. It harbors the fastest known outflow in a WF, with 22~GHz H$_2$O masers spanning a line-of-sight velocity range of 500~km~s$^{-1}$ \citep{gomez2011,gomez2015}. These masers trace two spatially separated clusters, with each having a high velocity dispersion of 170 km~s$^{-1}$. Through the trigonometric parallax measurement of its H$_2$O masers, I18113 is estimated to be at a distance of $\simeq$12~kpc (Orosz et al., in preparation), implying that 1~mas corresponds to a linear size of $\simeq$12~au. In this paper, we present the fine-scale structure and internal motions of 22~GHz H$_2$O masers around I18113, derived from multi-epoch VLBI observations.

\section{Observations and data reduction}
\label{sec:observations}

We observed the H$_2$O $6_{16}$$\rightarrow$$5_{23}$ maser line, with a rest frequency of 22.235080~GHz, towards I18113 using the Very Long Baseline Array (VLBA). Observations were carried out over five epochs on 2014 December 11, December 24, and 2015 January 5, January 19, and February 2 under project code BG231. The short epoch spacing ensured successful cross-identification of individual maser features across several epochs, facilitating proper motion measurements.

Each session lasted $\sim$5 hours with a target on-source time of $\sim$2.5 hours. Every hour the bright continuum sources NRAO530 (J1733$-$1304) or J1743$-$0350 were scanned for 4--5 minutes for delay and bandpass corrections. The observed velocity range was 860~km~s$^{-1}$, from $-$330~km~s$^{-1}$ to 530~km~s$^{-1}$ in V$_{\text{LSR}}$, where LSR denotes the kinematic definition of the local standard of rest. Recorded data were correlated by NRAO with VLBA-DiFX \citep{deller2011}. The single 64\,MHz-wide band was split into 2048 channels with a spacing of 31.25~kHz (0.42~km~s$^{-1}$ at 22~GHz).

We reduced the data using the NRAO AIPS package with a standard approach. After amplitude, delay, and bandpass corrections, we applied phase solutions obtained through fringe fitting and self-calibration from a reference maser channel (V$_{\text{LSR}}$=$-$148.2~km~s$^{-1}$) to all other channels. Due to self-calibration, only phase terms describing the relative offsets from the maser spot in the reference channel were measured. Image cubes were made using CLEAN deconvolution, with typical synthesized beams of 1.4$\times$0.3~mas. Maser detections were cataloged at a signal-to-noise cutoff of 7 by fitting Gaussian models to emission peaks \citep{imai2013pasj}. We grouped maser spots within $\sim$1 mas (our angular resolution), in each spectral feature, into maser features. The feature position was measured as the flux-density-weighted average position of each group of spots, corrected for observational and processing position offsets between epochs \citep{burns2017}.

\section{Results}
\label{sec:results}

We identified 130--170 maser features at each individual epoch of observation, out of which 92 features could be explicitly traced (based on their spatio-kinematic proximity) in all, and 55 additional features in at least three consecutive epochs. Our analysis uses these 147 (92+55) features (Table~\ref{table:masers-sample}). Image cube coregistration was done by setting the geometric center (error$\approx$0.005~mas) of the maser features identified in all epochs as a common reference point, assuming point symmetry \citep[e.g.,][]{chong2015}.

%TABLE:MEASUREMENTS
\begin{table*}
\caption{\label{table:masers-sample}Parameters of the 22~GHz H$_2$O maser features around IRAS~18113$-$2503 detected with the VLBA.}
\begin{center}
\begin{tabular}{c c c c c c c c c c}
\hline
\hline
\noalign{\smallskip}
\multicolumn{1}{c}{Maser} & \multicolumn{1}{c}{LOS velocity} & \multicolumn{2}{c}{Position} & \multicolumn{2}{c}{Proper motion} & \multicolumn{2}{c}{3D velocity} & \multicolumn{1}{c}{Brightness} & \multicolumn{1}{c}{Structure} \\
\multicolumn{1}{c}{ID} & \multicolumn{1}{c}{$V_{\text{LSR}}$} & \multicolumn{1}{c}{$\Delta x$} & \multicolumn{1}{c}{$\Delta y$} & \multicolumn{1}{c}{$\mu_{x}$} & \multicolumn{1}{c}{$\mu_{y}$} & \multicolumn{1}{c}{$V_{\text{3D}}$} & \multicolumn{1}{c}{$i$} & \multicolumn{1}{c}{$I$} & \multicolumn{1}{c}{Arc ID} \\
\noalign{\smallskip}
 & (km~s$^{-1}$) & (mas) & (mas) & (mas~yr$^{-1}$) & (mas~yr$^{-1}$) & (km~s$^{-1}$) & (deg) & (Jy~beam$^{-1}$) & \\
\hline
\noalign{\smallskip}
1 & $-$154.0 & $-$12.26 & 70.39 &    0.02 & 2.03 & 273.7 & 65 &  0.33 & -- \\
2 & $-$152.5 & $-$8.89  & 70.64 &    0.09 & 1.26 & 256.9 & 74 &  0.09 & -- \\
3 & $-$148.2 & $-$19.15 & 47.24 & $-$0.15 & 1.68 & 260.7 & 68 & 29.40 & in \\
4 & $-$146.4 & $-$18.57 & 36.37 & $-$0.42 & 2.59 & 283.0 & 58 &  0.30 & in \\
5 & $-$144.8 & $-$19.28 & 37.20 & $-$0.30 & 0.72 & 243.0 & 79 &  0.79 & in \\
\hline
\end{tabular}\\
\end{center}
\justify
{\sc Note}---Positions are relative to $(\alpha,\delta)_{\text{J2000.0}}$=(18$^{\rm h}$14$^{\rm m}$26$\fs$70263,$-$25$\degr$02$\arcmin$55$\farcs$6981), derived from separate phase referencing observations (Orosz et al., in prep.), and refer to the first detection of a maser feature (at epochs 1--3). Coordinates and proper motions are in the eastward ($\Delta x$=$\Delta\alpha$cos$\delta$, $\mu_{x}$=$\mu_{\alpha}$cos$\delta$) and northward directions ($\Delta y$=$\Delta\delta$, $\mu_{y}$=$\mu_{\delta}$). Thermal noise dominated astrometric errors are calculated to be $<$0.07~mas, depending on maser brightness. Proper motion errors are derived from the linear least squares fitting with values of 0.01--0.20~mas~yr$^{-1}$, which are dependent on the number of epochs in the fitting and the astrometric errors of the maser features. Typical rms noise in the line-free channels of the image cubes is $\sim$5~mJy~beam$^{-1}$. Inclination angles $i$ are relative to the plane of the sky. Brightness $I$ and line-of-sight (LOS) V$_{\text{LSR}}$ values are the averages of all detections. Only maser features detected at least in three consecutive epochs are listed, with IDs $\leqslant$92 referring to features detected in all five. Features used for fitting the arcs are indicated in the last column. (This table is available in its entirety on the journal website in a machine-readable format.)
\end{table*}

\begin{figure*}
\centering
\includegraphics[width=1\textwidth]{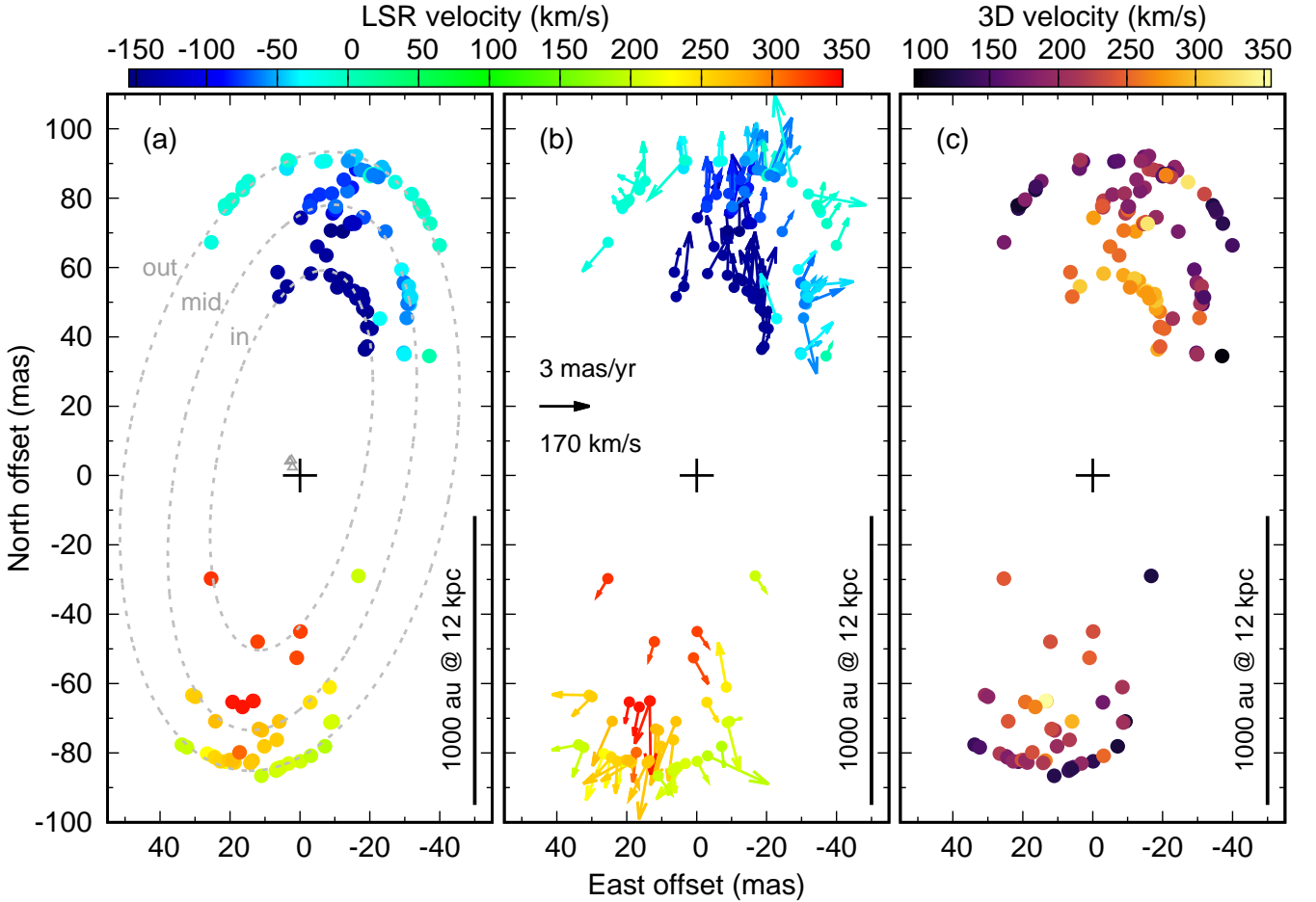}
\caption{(a) Spatial distribution, (b) internal proper motions, and (c) absolute values of the
three-dimensional velocities of the 22~GHz H$_2$O maser features around IRAS~18113$-$2503. Features in each lobe are found in distinct arcs, separated both spatially and in velocity. Grey dashed lines in panel (a) show ellipses fitted to each arc independently (`in', `mid', `out'; see Table~\ref{table:jets}), with grey triangles marking their centers. Map origin (black cross) is at the geometric center of the maser feature distribution and at the approximate location of the central object, at $(\alpha,\delta)_{\text{J2000.0}}$=(18$^{\rm h}$14$^{\rm m}$26$\fs$70263,$-$25$\degr$02$\arcmin$55$\farcs$6981). Shown coordinates are offsets relative to this position in the eastward ($\Delta\alpha$cos$\delta$) and northward directions ($\Delta\delta$). See the numerical values for these data in Table~\ref{table:masers-sample}.
}
\label{fig:masermap}
\end{figure*}

The VLBI maser maps are shown in Fig.~\ref{fig:masermap}. Maser features are found in two distinct clusters, separated along a polar geometric axis. The maximum extent of the maser region is $\sim$180~mas. The maser clusters are also widely separated in line-of-sight velocity, with the northern cluster clearly blueshifted with respect to the southern one. The systemic velocity of the host stellar system is unknown, but we assume V$_{\rm{LSR}}$ $\simeq$94~km~s$^{-1}$, equal to the center of the maser velocity range.

%FIG:MASER MAP AND PROPER MOTIONS

Most maser features trace three nested bipolar arcs (labeled `in', `mid', `out' in Fig.~\ref{fig:masermap}a) with a radially decreasing velocity gradient. The blueshifted northern region is generally brighter and more defined. Maser emission is also brighter in regions closer to the polar axis. The features in the bipolar arcs can be fitted with three independent ellipses that have a common center within 2.5~mas (and are coincident within errors, 1--5~mas), offset from the centroid of the maser feature distribution by $<$5~mas (Fig.~\ref{fig:masermap}a). Their major axes also share a common orientation, around PA$\simeq$170$\degr$, which are consistent within errors (2--5$\degr$). The spatial characteristics of the maser arcs indicate the presence of a bipolar collimated outflow, and the arcs are reminiscent of bow shocks at the tip of a jet \citep{boboltzmarvel2005}. The distinct pairs of arcs with nearly identical centers suggest an episodic jet-like phenomenon with a single origin.

We calculated the proper motion vectors of each maser feature by assuming linear motions and overall kinematic symmetry (Fig.~\ref{fig:masermap}b, Table~\ref{table:masers-sample}). The proper motions show that the arcs are expanding, and that the directions of the faster features are more parallel to each other. The absolute values of the 3D velocities in Fig.~\ref{fig:masermap}c (derived using the trigonometric distance; Sect.~\ref{sec:introduction}) reveal that the motions are fastest near the polar axis and slower away from it, along the arcs. The overall 3D velocities of the arcs decrease at larger separations from the central system, suggesting deceleration.

To understand the underlying phenomena of the inferred episodic and decelerating bow shocks, we derived their kinematic parameters (Table~\ref{table:jets}). Assuming that the star is at the center of the fitted ellipses shown in Fig.~\ref{fig:masermap}a, we characterize the average 3D velocity of all three shock fronts and their distances from the central star. The spatio-kinematic model presumes that the bow shocks share a common inclination (i=55$\degr$ with respect to the plane of the sky, the average inclination of all maser features). In reality the inclination angles of the individual maser features vary with $\sigma$$\approx$11$\degr$, but there is no clear systematic trend in their values. In addition, it is reasonable to expect that the bow shocks share a common inclination, as their orientation on the sky is also nearly identical.

%TABLE:JET PARAMETERS
\begin{table}
\caption{\label{table:jets} Derived parameters of the observed 22~GHz H$_2$O maser outflow around IRAS~18113$-$2503.}
\begin{center}
\begin{tabular}{r c c c c}
\hline
\hline
Arc & Length~($x$) & $V_{\text{3D}}$ & $t\,[v$=$\text{const}]$ & $t\,[v$=$v_0 e^{-kx}]$ \\
\noalign{\smallskip}
ID & (au) & (km~s$^{-1}$) & (yr) & (yr) \\
\hline
\noalign{\smallskip}
in  & 1160$\pm$25 & 280$\pm$10 & 20$\pm$2 & 12$\pm$2 \\
mid & 1610$\pm$10 & 180$\pm$15 & 42$\pm$4 & 21$\pm$4 \\
out & 1890$\pm$10 & 140$\pm$15 & 64$\pm$6 & 30$\pm$6 \\
\hline
\end{tabular}\\
\end{center}
{\sc Note}---IDs refer to labels in Fig.~\ref{fig:masermap}a. Length is the deprojected semimajor axis of the ellipses. $V_{\text{3D}}$ values are from averaging the velocities of the maser features used for fitting the individual ellipses (using only the blueshifted cluster). A common inclination of 55$\degr$ is adopted for all three bow shocks. The last two columns give the kinematic ages of the jets, by assuming outflow episodes with constant but different velocities of $v_0$=280, 180, 140~km~s$^{-1}$ ($t\,[v$=$\text{const}]$), or episodes with the same ejection velocity of $v_0$=840~km~s$^{-1}$ and a common exponential deceleration ($t\,[v$=$v_0 e^{-kx}]$); see Sect.~\ref{sec:discussion}.
\end{table}

\section{Discussion}
\label{sec:discussion}

Originally, \citet{gomez2011} proposed several possibilities to explain the large velocity dispersion of the I18113 H$_2$O maser clusters. The scenarios included projection effects due to the large opening angle or precessing motion of the outflow, and intrinsic characteristics in the outflow itself such as shocks produced by episodic mass ejections. Hints of episodic behavior have also been observed in other WFs \citep{chong2015,yung2011,sahai2017}. The case of I18113 is the first where we find multiple bow shock structures in a WF traced by masers, raising the unique possibility to derive 3D kinematic parameters at multiple points along the outflow. The velocity gradients of the maser features both along and perpendicular to the outflow axis (Fig.~\ref{fig:masermap}c) are not consistent with projection effects alone, suggesting instead interactions with an ambient medium \citep[e.g.,][]{raga1990,lee2001}.

A physical interpretation could be that of a single outflow permeating a stratified CSE produced by previous mass loss episodes in the AGB phase. In this case, masers trace shocks generated at each shell layer as the outflow proceeds outward \citep{kwok2008,chong2015}. While we cannot discard this scenario with the presently available data, the observed velocity and spatial signatures of the bow shocks are more naturally explained with collimated, episodic mass ejections. We can consider two basic scenarios: in the first, the initial ejection velocities increase each time; in the second, the ejection velocities remain the same, but the ejected mass clumps decelerate \citep[e.g.,][]{raga1990}.

\begin{figure*}
\centering
\includegraphics[width=1\textwidth,trim=0mm 0mm 0mm 40mm,clip]{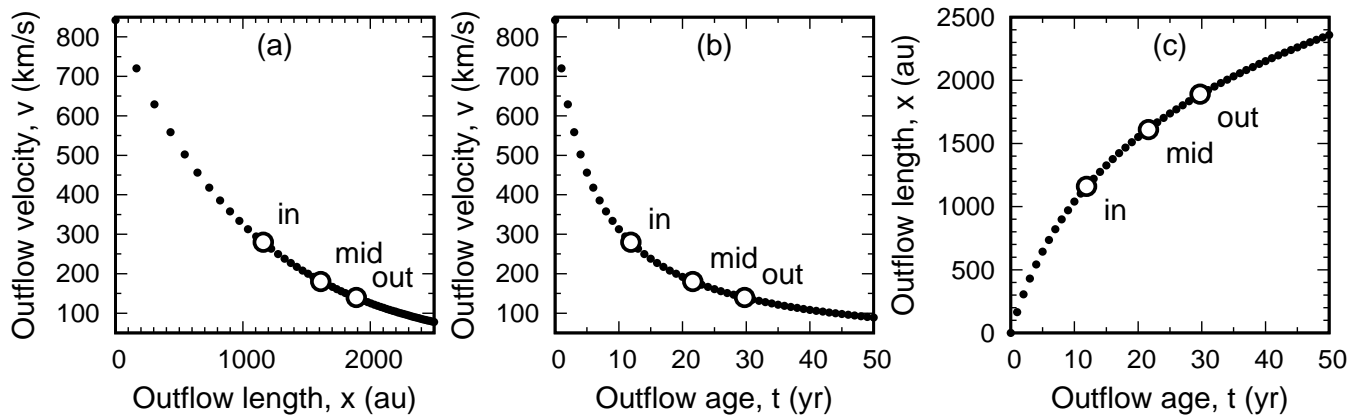}
\vspace{-2mm}
\caption{A fitted outflow model describing the motions of the masers, plotted as (a) jet velocity as a function of its spatial length; (b) jet velocity as a function of age; and (c) the spatial length of the jet as a function of its age. Open circles mark the observed values, while filled circles show the model for every whole year in time. Labels refer to Table~\ref{table:jets}.}
\label{fig:jetmodel}
\end{figure*}

The observed decrease in the 3D velocities of the three ejections is too violent to be explained with only a single steady outflow traveling through a uniform medium, which would imply the outflow slowing down linearly with distance \citep{canto2006}. Instead, the deceleration is clearly non-linear, and it can be well described with a common exponential decay as a function of distance $x$ from the source, $v(x)$=$v_{0} e^{-kx}$, supporting a decelerating outflow scenario. Fitting $v(x)$ to our three position--velocity points, we obtain the parameters $v_{0}$=$840\pm150$~km~s$^{-1}$ and $k$=$(6.4\pm0.9) \cdot 10^{-17}$~cm$^{-1}$ (Table~\ref{table:jets}, Fig.~\ref{fig:jetmodel}a). In such a case, each ejection produces bow shocks that then propagate outward in the ambient medium and decelerate as they interact with it.

%FIG:JET MODEL

Since the observed shocks are highly supersonic (Mach$>$70, with $v_{\text{sound}}$=2~km~s$^{-1}$ at 100~K), the mass ejections have high Reynolds numbers($\gtrsim$10$^{6}$), i.e. the outflow is expected to be turbulent. Under such conditions, the motion of the outflow can be modeled with the quadratic drag equation, $F_{\text{D}}$=$\rho C_{\text{D}} A v^{2}/2$, that describes the resistive force resulting from the outflow moving through an ambient medium. Here, $\rho$ is the mass density of the CSE, $C_{\text{D}}$ is the drag coefficient of the outflowing mass clumps, $A$ is the cross sectional area of a clump, and $v$ is the outflow velocity relative to the ambient medium. Taking $k$=$\rho C_{\text{D}} A /(2m)$ as a constant (where $m$ is the mass of the individual outflowing clumps), we can derive equations that describe the motion of the outflow and its evolution with time under the influence of drag, i.e. $v(x)$=$v_{0} e^{-kx}$, $v(t)$=$v_{0} (1 + v_{0} kt)^{-1}$, and $x(t)$=$\ln(1 + v_{0} kt) k^{-1}$. By adopting the drag scenario, we can reproduce the observed exponential decrease in the outflow velocity $v(x)$ with distance $x$ from the central source, and give physical meaning to the parameters $v_{0}$ and $k$ that were derived from our curve fitting with the values mentioned earlier (Fig.~\ref{fig:jetmodel}a). Using the above equations, we can also describe the velocity and position of the outflowing mass as a function of time, $v(t)$ and $x(t)$, respectively (Figs.~\ref{fig:jetmodel}b and \ref{fig:jetmodel}c), and calculate the ages of the different ejections (Table~\ref{table:jets}).

The kinematic ages derived assuming outflows with constant but different velocities $v$ = $V_{\text{3D}}$ ($t\,[v$=$\text{const}]$), and those derived from the exponential decrease in velocity (t\,[$v$=$v_0 e^{-kx}]$ with $v_{0}$=840~km~s$^{-1}$), differ only by a factor of two (Table~\ref{table:jets}). Both values depend on initial assumptions, e.g., we neglected the ambient wind velocity of the CSE, assumed that the CSE has a constant density within the region encompassed by the maser emission, and took the star to be halfway between the maser bow shocks both spatially and in velocity. However, the relative age differences between the episodes ($\Delta t$) are equal within errors regardless of the adopted ejection scenario and only their absolute values change, i.e., with or without considering drag we reach a similar conclusion; that the jet-like ejections from I18113 occur periodically, every $\sim$20~yr vs $\sim$10~yr (Table~\ref{table:jets}). Although the second scenario predicts that a new ejection could have been launched by the time of these observations, its non-detection is reasonable, since maser emission is very sensitive to physical conditions in the medium and the newest mass-loss event may not have yet reached the maser pumping region. Therefore, monitoring the evolution of the H$_2$O masers is essential as the detection of a new ejection could test our predictions.

Our kinematic model is just a qualitative first approach, as the drag equation only describes the motion of a supersonic outflow through an enclosing `viscous fluid', i.e. the relic CSE, and it is not a hydrodynamic simulation describing shock formation and propagation. Furthermore, our model assumes a constant density in the ambient medium, whereas in reality the first ejection could sweep up the CSE material, reducing the density of the medium and allowing the following ejections to propagate through the sparser environment faster than assumed in the model \citep[e.g.,][]{krist2008}. In this case, the parameter $v_{0}$ would not be necessarily the initial ejection velocity. Nevertheless, despite its simplicity, the kinematic model reproduces well the non-linear deceleration in the data.

Using our derived equations, if we follow the trajectory of a single maser cloud, we can estimate the density of the ambient medium of the CSE that is slowing down the ejections. Assuming that each maser feature is a spherical cloud tracing post-shock gas in a turbulent flow ($C_{\text{D}}$=0.47), with a diameter of 1~au \citep[e.g.,][]{hollenbach2013}, and density $n_{\text{H}_2}$(maser)=$10^{9}$~cm$^{-3}$ \citep[e.g.,][]{elitzur1989}, it is possible to derive from the constant $k$ the number density of the ambient medium to be $n_{\text{H}_2}$$\simeq$$3\cdot 10^{6}$~cm$^{-3}$. Due to the many simplifying assumptions, this result should only be taken as an order of magnitude estimate. In particular, the CSE might have a radially isotropic density profile ($\rho \propto r^{-2}$, where $r$ is the CSE radius), so a constant density assumption introduces an uncertainty of $\sim$30 per cent. Maser clouds can also be flattened into cylindrical slabs in shocked regions \citep{richards2011,hollenbach2013}, which could increase $A$ and $C_{\text{D}}$ by a factor of a few.

The derived ambient density implies that within a sphere of 2000~au radius (the extent of the maser region) there is about 0.5~M$_\odot$ in the CSE (with an uncertainty of $\sim$30 per cent), corresponding to an average mass loss rate in the range of $3\cdot 10^{-4}$ to $1.4\cdot 10^{-5}$~M$_\odot$~yr$^{-1}$ for typical AGB wind speeds of 10--20~km~s$^{-1}$. These are reasonable numbers for a star with an initial mass of 4--8~M$_\odot$, which is the range where the progenitors of WFs are believed to be, based on e.g., CO observations \citep{he2008,imai2012HBB,rizzo2013}, optical extinction \citep{suarez2008}, and the Galactic scale heights of WFs \citep{imai+2007}.

We interpret the observed periodicity to come from an accreting binary system that would also explain the collimation of the outflow, as predicted by various stellar binary models \citep{garciafrank2004,soker2000,soker2015,nordhaus2006}. Stars with an initial mass of 4--8~M$_\odot$ will evolve into remnant central stars of $\sim$0.8--1~M$_\odot$ after the AGB \citep{kalirai2008,renedo2010}. A central mass in this range, together with a binary period of $\sim$10~yr (from the observed periodicity) would result in a binary separation on the order of 10~au. Among binary central stars in PNe, very few are known with decade-long orbital periods \citep{miszalski2018}. This means that maser observations might help us identify long period central binaries, which are difficult to detect with other observational techniques. Our scenario to explain the maser spatio-kinematics (Fig.~\ref{fig:jetmodel}) requires a dense and mostly unperturbed CSE along the jet axis. However, a binary system with a separation of only $\sim$10~au may go through a wind Roche-lobe overflow phase \citep{mohamed2012} during the AGB expansion of the primary. This could affect the envelope geometry by creating expanding toroids with low density along the poles, as observed in other bipolar post-AGB stars \citep[e.g.,][]{sahai2006,sahai2017,olofsson2015,gomez2018}, unless the companion is of very low-mass. Thus, considering our estimated high gas density along the jet axis, we speculate that the binary companion could be of low mass (even substellar).

In conclusion, VLBI maser observations reveal, for the first time, an exponentially decelerating system of episodic ejections from a WF, exhibiting a periodicity of $\sim$10~yr. Presently, the observations may best be explained by a close ($\sim$10~au) binary. Future observations of thermal molecular lines are necessary to disclose the circumstellar structure and the properties of the outflow in this WF, providing more information on the inferred binary central star.

\section*{Acknowledgements}

We thank Jorge Cant\'{o} for the helpful discussion on outflow kinematics and our referee, Anita Richards, for a constructive peer review.
GO was supported by the Xinjiang Key Laboratory of Radio Astrophysics grant 2016D03020, NSFC 11503072, the National Key R\&D Program of China grant 2018YFA0404602, the Youth Innovation Promotion Association of the Chinese Academy of Sciences, the Joint Institute of VLBI ERIC, Kagoshima University, and the MEXT Monbukagakusho scholarship.
JFG, JMT, and LFM are partially supported by MINECO (Spain) grants AYA2014-57369-C3 and AYA2017-84390-C2-R (co-funded by FEDER).
HI and GO were supported by the MEXT KAKENHI program (16H02167).
HI, JMT and JFG were supported by the Invitation Program for Foreign Researchers of the Japan Society for Promotion of Science (JSPS grants S17115 and S14128).
DT was supported by the ERC consolidator grant 614264.
MAG acknowledges support from the MINECO grant AYA2014-57280-P, co-funded by FEDER.
MAPT received support from the MINECO grants AYA2012-38491-C02-02 and AYA2015-63939-C2-1-P.
JRR acknowledges support from the project ESP2015-65597-C4-1-R (MINECO and FEDER).
LU received support from the University of Guanajuato (Mexico) grant ID CIIC 17/2018.
The VLBA is run by the Long Baseline Observatory, a National Science Foundation facility operated under cooperative agreement by Associated Universities, Inc. We note the use of the Astrogeo Center.

%%%%%%%%%%%%%%%%%%%%%%%%%%%%%%%%%%%%%%%%%%%%%%%%%%

%%%%%%%%%%%%%%%%%%%% REFERENCES %%%%%%%%%%%%%%%%%%

%%%%%%%%%%%%%%%%%%%%%%%%%%%%%%%%%%%%%%%%%%%%%%%%%%

% Don't change these lines
\bsp	% typesetting comment
\label{lastpage}

%%%%%%%%%%%%%%%%%%%%%%%%%%%%%%%%%%%%%%%%%%%%%%%%%%

%Only-only material
\appendix
\setcounter{table}{0}
\renewcommand{\thetable}{A\arabic{table}}

%TABLE:MEASUREMENTS
\begin{table*}
\caption{\label{table:masers-online}Parameters of the 22~GHz H$_2$O maser features around IRAS~18113$-$2503 detected with the VLBA. (Online supplementary material.)}
\begin{center}
\begin{tabular}{c c c c c c c c c c}
\hline
\hline
\noalign{\smallskip}
\multicolumn{1}{c}{Maser} & \multicolumn{1}{c}{LOS velocity} & \multicolumn{2}{c}{Position} & \multicolumn{2}{c}{Proper motion} & \multicolumn{2}{c}{3D velocity} & \multicolumn{1}{c}{Brightness} & \multicolumn{1}{c}{Structure} \\
\multicolumn{1}{c}{ID} & \multicolumn{1}{c}{$V_{\text{LSR}}$} & \multicolumn{1}{c}{$\Delta x$} & \multicolumn{1}{c}{$\Delta y$} & \multicolumn{1}{c}{$\mu_{x}$} & \multicolumn{1}{c}{$\mu_{y}$} & \multicolumn{1}{c}{$V_{\text{3D}}$} & \multicolumn{1}{c}{$i$} & \multicolumn{1}{c}{$I$} & \multicolumn{1}{c}{Arc ID} \\
\noalign{\smallskip}
 & (km~s$^{-1}$) & (mas) & (mas) & (mas~yr$^{-1}$) & (mas~yr$^{-1}$) & (km~s$^{-1}$) & (deg) & (Jy~beam$^{-1}$) & \\
\hline
\noalign{\smallskip}
1 & $-$154.0 & $-$12.26 & 70.39 &    0.02 & 2.03 & 273.7 & 65 &  0.33 & -- \\
2 & $-$152.5 & $-$8.89  & 70.64 &    0.09 & 1.26 & 256.9 & 74 &  0.09 & -- \\
3 & $-$148.2 & $-$19.15 & 47.24 & $-$0.15 & 1.68 & 260.7 & 68 & 29.40 & in \\
4 & $-$146.4 & $-$18.57 & 36.37 & $-$0.42 & 2.59 & 283.0 & 58 &  0.30 & in \\
5 & $-$144.8 & $-$19.28 & 37.20 & $-$0.30 & 0.72 & 243.0 & 79 &  0.79 & in \\
6 & $-$144.8 & $-$13.15 & 56.15 & $-$0.53 & 2.52 & 280.4 & 58 & 0.47 & in \\
7 & $-$144.2 & $-$18.14 & 48.20 & $-$0.26 & 2.56 & 279.6 & 58 & 10.57 & in \\
8 & $-$143.6 & 3.63 & 54.51 & $-$0.34 & 3.28 & 302.7 & 52 & 0.43 & in \\
9 & $-$142.7 & $-$8.77 & 57.80 & $-$0.10 & 2.68 & 281.8 & 57 & 12.33 & in \\
10 & $-$141.8 & $-$17.97 & 50.40 & 0.27 & 3.30 & 301.8 & 51 & 0.62 & in \\
11 & $-$140.2 & $-$14.39 & 53.30 & $-$0.48 & 2.60 & 278.5 & 57 & 11.71 & in \\
12 & $-$137.6 & $-$18.12 & 50.56 & $-$0.95 & 3.19 & 299.2 & 51 & 0.34 & in \\
13 & $-$134.2 & $-$0.23 & 74.33 & $-$1.39 & 2.33 & 275.7 & 56 & 0.33 & -- \\
14 & $-$129.8 & $-$14.34 & 72.76 & 0.43 & 2.79 & 275.5 & 54 & 0.97 & -- \\
15 & $-$126.4 & $-$14.61 & 72.58 & $-$0.13 & 0.57 & 223.0 & 81 & 0.12 & -- \\
16 & $-$123.9 & $-$15.10 & 73.06 & 0.21 & 1.56 & 235.7 & 68 & 0.16 & -- \\
17 & $-$112.3 & $-$9.40 & 75.59 & $-$0.20 & 1.21 & 217.9 & 71 & 0.43 & -- \\
18 & $-$92.4 & $-$10.19 & 76.51 & $-$0.42 & 1.88 & 216.3 & 60 & 0.17 & -- \\
19 & $-$90.1 & $-$10.17 & 76.58 & $-$0.10 & 1.71 & 208.3 & 62 & 0.65 & -- \\
20 & $-$86.7 & $-$16.14 & 88.29 & $-$0.33 & 2.28 & 223.5 & 54 & 0.08 & -- \\
21 & $-$84.9 & $-$14.82 & 82.99 & $-$0.15 & 1.26 & 193.1 & 68 & 0.17 & -- \\
22 & $-$81.7 & $-$2.91 & 77.38 & $-$0.83 & 2.18 & 220.3 & 53 & 0.10 & -- \\
23 & $-$80.4 & $-$3.14 & 79.25 & $-$0.17 & 2.34 & 219.7 & 53 & 0.14 & -- \\
24 & $-$79.2 & $-$5.66 & 81.03 & 0.15 & 1.24 & 187.3 & 68 & 0.46 & -- \\
25 & $-$77.0 & $-$9.17 & 81.32 & $-$0.99 & 1.62 & 202.3 & 58 & 1.08 & -- \\
26 & $-$74.1 & $-$18.01 & 74.49 & $-$0.37 & 1.49 & 189.5 & 63 & 0.09 & mid \\
27 & $-$72.3 & $-$18.21 & 74.50 & $-$0.24 & 1.78 & 195.3 & 58 & 0.31 & mid \\
28 & $-$71.5 & $-$10.14 & 77.22 & 0.18 & 2.26 & 209.7 & 52 & 0.17 & mid \\
29 & $-$67.5 & $-$10.02 & 77.51 & $-$0.28 & 2.07 & 200.6 & 54 & 0.35 & mid \\
30 & $-$67.2 & $-$29.82 & 55.34 & $-$0.78 & 1.78 & 195.6 & 56 & 0.54 & mid \\
31 & $-$64.5 & $-$24.48 & 70.37 & $-$0.60 & 1.08 & 173.4 & 66 & 0.15 & mid \\
32 & $-$64.2 & $-$13.61 & 82.01 & $-$0.37 & 1.28 & 175.6 & 64 & 0.11 & -- \\
33 & $-$62.8 & $-$17.43 & 88.14 & $-$0.54 & 2.33 & 207.8 & 49 & 2.97 & -- \\
34 & $-$61.5 & $-$18.89 & 87.83 & 0.06 & 2.47 & 209.6 & 48 & 4.92 & -- \\
35 & $-$59.7 & $-$17.35 & 88.22 & 0.15 & 2.73 & 218.9 & 45 & 2.12 & -- \\
36 & $-$51.9 & $-$31.31 & 50.93 & $-$1.17 & 1.14 & 173.2 & 57 & 0.08 & mid \\
37 & $-$51.5 & $-$24.09 & 87.88 & $-$1.15 & 1.49 & 180.8 & 54 & 0.18 & -- \\
38 & $-$50.8 & $-$31.20 & 52.23 & $-$0.53 & 0.96 & 157.8 & 67 & 0.08 & mid \\
39 & $-$49.9 & $-$31.38 & 49.41 & $-$0.71 & 0.43 & 151.6 & 72 & 0.08 & mid \\
40 & $-$48.0 & $-$23.48 & 88.98 & $-$0.43 & 2.08 & 186.7 & 50 & 0.08 & -- \\
41 & $-$48.0 & $-$31.44 & 51.00 & $-$0.83 & 1.19 & 164.3 & 60 & 0.07 & mid \\
42 & $-$46.7 & $-$14.64 & 91.88 & $-$0.93 & 1.04 & 161.6 & 61 & 0.15 & -- \\
43 & $-$44.9 & $-$29.69 & 35.38 & $-$1.27 & 0.69 & 161.5 & 59 & 3.75 & -- \\
44 & $-$44.3 & $-$15.09 & 91.19 & 0.15 & 1.97 & 178.3 & 51 & 0.20 & -- \\
45 & $-$44.1 & $-$16.06 & 92.16 & $-$0.22 & 2.01 & 179.7 & 50 & 0.27 & -- \\
46 & $-$42.1 & $-$31.66 & 51.52 & $-$1.05 & 1.06 & 160.5 & 58 & 0.59 & mid \\
47 & $-$40.3 & $-$29.83 & 34.97 & $-$0.69 & 0.78 & 146.9 & 66 & 14.96 & -- \\
48 & $-$36.5 & 3.87 & 88.71 & 0.16 & 1.92 & 170.4 & 50 & 1.58 & -- \\
49 & $-$35.0 & $-$29.05 & 59.35 & $-$1.06 & 1.12 & 156.2 & 56 & 0.12 & mid \\
50 & $-$34.8 & $-$6.21 & 90.53 & $-$0.10 & 1.29 & 148.4 & 60 & 0.24 & -- \\
51 & $-$29.2 & 2.83 & 90.60 & 0.18 & 1.51 & 150.5 & 55 & 0.32 & out \\
52 & $-$25.0 & $-$35.42 & 75.90 & $-$0.93 & 1.09 & 144.2 & 56 & 0.13 & out \\
53 & $-$22.7 & $-$7.28 & 90.70 & 0.01 & 1.77 & 154.4 & 49 & 0.40 & out \\
54 & $-$21.7 & 21.25 & 77.02 & 0.59 & 0.75 & 128.0 & 65 & 0.24 & out \\
55 & $-$21.7 & $-$21.90 & 87.15 & $-$0.71 & 1.67 & 155.3 & 48 & 0.25 & out \\
56 & $-$17.3 & 25.36 & 67.23 & 1.53 & $-$1.74 & 172.5 & 40 & 0.06 & out \\
57 & $-$11.9 & $-$20.43 & 86.98 & $-$0.45 & 1.36 & 133.6 & 52 & 0.22 & out \\
58 & $-$10.0 & 19.56 & 79.07 & 0.53 & 0.79 & 117.3 & 63 & 0.11 & out \\
59 & $-$9.9 & 14.71 & 84.91 & 0.14 & 2.38 & 171.1 & 37 & 0.46 & out \\
\hline
\end{tabular}\\
\end{center}
\justify
{\sc Note}---Continued on the next page.
\end{table*}

\begin{table*}
\contcaption{Parameters of the 22~GHz H$_2$O maser features around IRAS~18113$-$2503 detected with the VLBA. (Online supplementary material.)}
\begin{center}
\begin{tabular}{c c c c c c c c c c}
\hline
\hline
\noalign{\smallskip}
\multicolumn{1}{c}{Maser} & \multicolumn{1}{c}{LOS velocity} & \multicolumn{2}{c}{Position} & \multicolumn{2}{c}{Proper motion} & \multicolumn{2}{c}{3D velocity} & \multicolumn{1}{c}{Brightness} & \multicolumn{1}{c}{Structure} \\
\multicolumn{1}{c}{ID} & \multicolumn{1}{c}{$V_{\text{LSR}}$} & \multicolumn{1}{c}{$\Delta x$} & \multicolumn{1}{c}{$\Delta y$} & \multicolumn{1}{c}{$\mu_{x}$} & \multicolumn{1}{c}{$\mu_{y}$} & \multicolumn{1}{c}{$V_{\text{3D}}$} & \multicolumn{1}{c}{$i$} & \multicolumn{1}{c}{$I$} & \multicolumn{1}{c}{Arc ID} \\
\noalign{\smallskip}
 & (km~s$^{-1}$) & (mas) & (mas) & (mas~yr$^{-1}$) & (mas~yr$^{-1}$) & (km~s$^{-1}$) & (deg) & (Jy~beam$^{-1}$) & \\
\hline
\noalign{\smallskip}
60 & $-$7.0 & 16.38 & 82.39 & 0.19 & 1.09 & 119.2 & 58 & 0.25 & out \\
61 & $-$5.8 & $-$39.93 & 66.39 & $-$0.73 & 1.64 & 142.9 & 44 & 0.28 & out \\
62 & $-$4.9 & $-$20.01 & 86.31 & 0.16 & 2.09 & 155.0 & 40 & 1.05 & out \\
63 & $-$3.2 & $-$37.22 & 72.68 & $-$0.78 & 1.22 & 127.5 & 50 & 0.19 & out \\
64 & $-$2.4 & $-$34.35 & 77.91 & $-$1.11 & 0.67 & 121.6 & 53 & 0.34 & out \\
65 & $-$1.5 & $-$34.98 & 76.78 & $-$0.31 & 1.63 & 134.4 & 45 & 0.41 & out \\
66 & 190.2 & $-$7.16 & $-$78.07 & $-$0.14 & $-$1.27 & 120.6 & $-$53 & 0.12 & out \\
67 & 200.2 & 33.76 & $-$77.61 & 1.60 & $-$0.32 & 140.9 & $-$49 & 0.34 & out \\
68 & 200.6 & 32.32 & $-$78.41 & 0.65 & $-$1.78 & 151.5 & $-$45 & 0.24 & out \\
69 & 201.6 & 6.84 & $-$85.08 & 0.28 & $-$1.49 & 137.8 & $-$51 & 0.23 & out \\
70 & 201.9 & $-$16.74 & $-$28.95 & $-$0.61 & $-$0.92 & 124.8 & $-$60 & 0.09 & -- \\
71 & 203.8 & $-$0.13 & $-$82.41 & $-$0.64 & $-$1.21 & 134.6 & $-$55 & 0.13 & out \\
72 & 207.9 & 5.47 & $-$84.11 & 0.40 & $-$1.59 & 146.9 & $-$51 & 1.88 & out \\
73 & 209.9 & 3.20 & $-$83.04 & 1.74 & $-$2.15 & 195.2 & $-$36 & 0.56 & out \\
74 & 211.7 & 6.47 & $-$84.48 & $-$0.09 & $-$1.47 & 144.3 & $-$55 & 5.05 & out \\
75 & 221.0 & 21.30 & $-$82.54 & 0.25 & $-$1.40 & 150.4 & $-$58 & 1.00 & out \\
76 & 226.5 & 24.71 & $-$80.96 & 0.30 & $-$1.90 & 171.7 & $-$50 & 0.32 & out \\
77 & 229.4 & 26.54 & $-$80.18 & 1.07 & $-$2.81 & 218.1 & $-$38 & 0.05 & out \\
78 & 253.8 & 24.39 & $-$81.28 & 1.00 & $-$1.15 & 181.7 & $-$62 & 5.87 & -- \\
79 & 255.5 & 30.77 & $-$63.35 & 2.22 & 0.05 & 205.1 & $-$52 & 0.82 & mid \\
80 & 256.5 & 22.71 & $-$82.40 & 0.10 & $-$1.95 & 196.8 & $-$56 & 0.92 & -- \\
81 & 259.3 & 18.67 & $-$82.78 & 1.81 & $-$1.11 & 204.6 & $-$54 & 0.24 & -- \\
82 & 260.5 & 29.92 & $-$63.79 & 1.19 & $-$1.18 & 191.6 & $-$60 & 1.91 & mid \\
83 & 261.6 & 10.17 & $-$78.07 & 0.25 & $-$1.96 & 201.5 & $-$56 & 0.48 & -- \\
84 & 263.5 & 20.18 & $-$82.06 & 2.55 & $-$1.42 & 237.3 & $-$46 & 3.42 & -- \\
85 & 264.6 & 13.53 & $-$82.19 & 0.61 & $-$3.52 & 265.5 & $-$40 & 0.18 & -- \\
86 & 267.1 & 10.89 & $-$73.52 & 0.05 & $-$2.24 & 214.7 & $-$54 & 0.26 & mid \\
87 & 269.4 & 24.15 & $-$70.85 & 1.18 & $-$2.69 & 242.0 & $-$46 & 0.12 & mid \\
88 & 322.1 & 0.88 & $-$52.56 & $-$0.88 & $-$1.58 & 250.1 & $-$66 & 2.11 & -- \\
89 & 324.2 & 12.11 & $-$47.93 & 0.33 & $-$1.12 & 239.4 & $-$74 & 0.16 & in \\
90 & 324.6 & $-$0.14 & $-$45.01 & $-$0.63 & $-$0.90 & 238.8 & $-$75 & 0.26 & in \\
91 & 330.7 & 25.39 & $-$29.73 & 0.65 & $-$1.06 & 246.9 & $-$73 & 0.10 & in \\
92 & 341.5 & 19.26 & $-$65.30 & 0.26 & $-$1.30 & 258.7 & $-$73 & 0.25 & -- \\
\noalign{\smallskip}
\noalign{\smallskip}
101 & $-$146.8 & $-$12.00 & 56.86 & 1.03 & 3.01 & 301.2 & 53 & 0.08 & in \\
102 & $-$145.8 & $-$15.48 & 53.23 & $-$0.25 & 3.55 & 313.9 & 50 & 0.10 & in \\
103 & $-$145.6 & $-$19.27 & 42.88 & 0.16 & 1.03 & 247.0 & 76 & 0.10 & in \\
104 & $-$145.3 & $-$10.71 & 54.25 & 0.09 & 2.03 & 266.0 & 64 & 0.06 & in \\
105 & $-$143.1 & $-$17.70 & 52.28 & 1.13 & 2.63 & 287.8 & 56 & 0.17 & in \\
106 & $-$142.6 & $-$20.43 & 42.35 & $-$0.13 & 1.32 & 248.5 & 72 & 0.25 & in \\
107 & $-$142.6 & $-$16.23 & 51.05 & $-$0.54 & 2.39 & 274.7 & 60 & 0.30 & in \\
108 & $-$140.7 & $-$3.13 & 58.21 & $-$1.49 & 2.70 & 293.3 & 53 & 0.08 & in \\
109 & $-$139.7 & $-$7.56 & 63.47 & $-$0.59 & 1.44 & 250.0 & 69 & 0.16 & -- \\
110 & $-$139.3 & 6.38 & 58.64 & $-$0.33 & 1.44 & 248.1 & 70 & 0.19 & -- \\
111 & $-$139.3 & 5.86 & 51.58 & $-$0.63 & 1.38 & 248.9 & 70 & 0.06 & in \\
112 & $-$138.8 & $-$15.43 & 52.71 & 0.19 & 2.51 & 273.4 & 58 & 0.55 & in \\
113 & $-$135.1 & $-$4.99 & 66.04 & 0.44 & 1.85 & 253.5 & 65 & 0.05 & -- \\
114 & $-$121.2 & $-$15.69 & 72.63 & 1.17 & 4.16 & 326.8 & 41 & 0.58 & -- \\
115 & $-$84.6 & $-$10.20 & 76.59 & 0.02 & 3.00 & 247.2 & 46 & 0.10 & -- \\
116 & $-$78.2 & $-$12.55 & 85.30 & $-$0.24 & 1.10 & 183.7 & 70 & 0.09 & -- \\
117 & $-$76.8 & $-$2.79 & 77.62 & $-$0.06 & 1.60 & 193.7 & 62 & 0.17 & mid \\
118 & $-$75.1 & $-$2.75 & 77.72 & $-$0.03 & 2.99 & 240.0 & 45 & 0.10 & mid \\
119 & $-$65.4 & $-$29.77 & 55.69 & $-$0.74 & 1.45 & 184.5 & 60 & 0.20 & mid \\
120 & $-$65.1 & $-$10.25 & 77.60 & 0.09 & 0.33 & 160.4 & 83 & 0.08 & mid \\
121 & $-$63.9 & $-$22.34 & 86.06 & $-$1.01 & 2.55 & 222.1 & 45 & 0.67 & -- \\
122 & $-$63.9 & $-$10.28 & 77.51 & $-$1.14 & 0.87 & 177.8 & 63 & 0.15 & mid \\
123 & $-$62.7 & $-$30.83 & 49.57 & $-$2.19 & 0.86 & 206.3 & 49 & 0.14 & mid \\
124 & $-$62.3 & $-$20.98 & 86.64 & $-$1.30 & 3.45 & 261.7 & 37 & 0.60 & -- \\
125 & $-$59.2 & $-$30.56 & 45.45 & $-$0.77 & $-$3.36 & 248.8 & 38 & 0.06 & mid \\
\hline
\end{tabular}\\
\end{center}
\justify
{\sc Note}---Continued on the next page.
\end{table*}

\begin{table*}
\contcaption{Parameters of the 22~GHz H$_2$O maser features around IRAS~18113$-$2503 detected with the VLBA. (Online supplementary material.)}
\begin{center}
\begin{tabular}{c c c c c c c c c c}
\hline
\hline
\noalign{\smallskip}
\multicolumn{1}{c}{Maser} & \multicolumn{1}{c}{LOS velocity} & \multicolumn{2}{c}{Position} & \multicolumn{2}{c}{Proper motion} & \multicolumn{2}{c}{3D velocity} & \multicolumn{1}{c}{Brightness} & \multicolumn{1}{c}{Structure} \\
\multicolumn{1}{c}{ID} & \multicolumn{1}{c}{$V_{\text{LSR}}$} & \multicolumn{1}{c}{$\Delta x$} & \multicolumn{1}{c}{$\Delta y$} & \multicolumn{1}{c}{$\mu_{x}$} & \multicolumn{1}{c}{$\mu_{y}$} & \multicolumn{1}{c}{$V_{\text{3D}}$} & \multicolumn{1}{c}{$i$} & \multicolumn{1}{c}{$I$} & \multicolumn{1}{c}{Arc ID} \\
\noalign{\smallskip}
 & (km~s$^{-1}$) & (mas) & (mas) & (mas~yr$^{-1}$) & (mas~yr$^{-1}$) & (km~s$^{-1}$) & (deg) & (Jy~beam$^{-1}$) & \\
\hline
\noalign{\smallskip}
126 & $-$56.3 & $-$13.78 & 90.73 & $-$0.70 & 1.94 & 190.6 & 52 & 0.12 & -- \\
127 & $-$41.0 & $-$31.12 & 54.63 & $-$2.86 & 0.26 & 211.9 & 40 & 0.07 & mid \\
128 & $-$40.3 & $-$31.77 & 51.39 & $-$0.79 & 0.57 & 145.4 & 68 & 0.57 & mid \\
129 & $-$39.4 & 3.75 & 88.48 & 0.32 & 1.61 & 163.0 & 55 & 0.14 & -- \\
130 & $-$36.1 & $-$29.85 & 34.98 & $-$1.90 & 1.92 & 201.5 & 40 & 0.07 & -- \\
131 & $-$30.7 & $-$22.88 & 45.22 & 0.96 & 2.71 & 205.9 & 37 & 0.07 & -- \\
132 & $-$18.8 & $-$27.26 & 84.64 & 1.03 & 5.23 & 323.8 & 20 & 0.06 & out \\
133 & $-$17.5 & 16.20 & 83.31 & 1.33 & 0.69 & 140.4 & 53 & 0.07 & out \\
134 & $-$14.0 & $-$31.95 & 81.15 & $-$3.44 & $-$0.66 & 226.8 & 28 & 0.05 & out \\
135 & $-$12.1 & 21.44 & 77.90 & 0.54 & 0.23 & 111.3 & 73 & 0.07 & out \\
136 & $-$11.8 & 20.44 & 78.32 & 0.31 & $-$0.15 & 107.6 & 80 & 0.05 & out \\
137 & $-$9.9 & 3.53 & 90.97 & 2.11 & $-$2.33 & 207.1 & 30 & 0.06 & out \\
138 & $-$8.6 & 19.30 & 79.57 & $-$0.92 & 2.44 & 180.4 & 35 & 0.07 & out \\
139 & 8.6 & $-$36.98 & 34.44 & $-$0.31 & 0.70 & 95.9 & $-$63 & 0.05 & -- \\
140 & 194.7 & $-$3.11 & $-$80.85 & $-$3.65 & $-$1.70 & 250.0 & $-$24 & 0.04 & out \\
141 & 199.0 & 11.00 & $-$86.59 & $-$0.78 & $-$1.10 & 130.1 & $-$54 & 0.08 & out \\
142 & 202.4 & $-$9.34 & $-$70.93 & $-$0.76 & 0.21 & 117.1 & $-$68 & 0.07 & out \\
143 & 204.9 & 6.09 & $-$84.58 & 0.38 & $-$1.20 & 131.9 & $-$57 & 1.74 & out \\
144 & 207.9 & $-$8.73 & $-$71.24 & $-$0.73 & $-$2.91 & 205.1 & $-$34 & 0.08 & out \\
145 & 239.3 & $-$8.46 & $-$61.01 & 0.48 & 2.68 & 212.4 & $-$43 & 0.07 & mid \\
146 & 249.2 & $-$2.92 & $-$65.40 & $-$0.67 & $-$1.19 & 173.5 & $-$63 & 0.06 & mid \\
147 & 261.4 & 6.64 & $-$76.21 & 0.30 & $-$2.47 & 219.1 & $-$50 & 0.13 & -- \\
148 & 262.7 & 14.17 & $-$82.91 & 1.07 & $-$2.20 & 218.6 & $-$50 & 0.09 & -- \\
149 & 265.6 & 5.90 & $-$70.91 & 1.54 & $-$3.83 & 290.9 & $-$36 & 0.06 & -- \\
150 & 268.3 & 19.19 & $-$82.10 & $-$0.53 & $-$1.61 & 199.1 & $-$61 & 0.16 & -- \\
151 & 269.8 & 11.75 & $-$73.10 & 0.08 & $-$2.70 & 233.4 & $-$49 & 0.11 & mid \\
152 & 308.9 & 17.25 & $-$79.85 & 0.22 & $-$1.79 & 238.0 & $-$64 & 0.06 & -- \\
153 & 337.7 & 13.30 & $-$65.06 & 0.91 & $-$2.56 & 288.3 & $-$58 & 0.18 & mid \\
154 & 339.9 & 13.42 & $-$65.00 & $-$0.06 & $-$4.43 & 351.9 & $-$44 & 0.11 & -- \\
155 & 342.3 & 16.40 & $-$66.75 & 0.33 & $-$1.73 & 267.5 & $-$68 & 0.12 & -- \\
\hline
\end{tabular}\\
\end{center}
\justify
{\sc Note}---Positions are relative to $(\alpha,\delta)_{\text{J2000.0}}$=(18$^{\rm h}$14$^{\rm m}$26$\fs$70263,$-$25$\degr$02$\arcmin$55$\farcs$6981), derived from separate phase referencing observations (Orosz et al., in prep.), and refer to the first detection of a maser feature (at epochs 1--3). Coordinates and proper motions are in the eastward ($\Delta x$=$\Delta\alpha$cos$\delta$, $\mu_{x}$=$\mu_{\alpha}$cos$\delta$) and northward directions ($\Delta y$=$\Delta\delta$, $\mu_{y}$=$\mu_{\delta}$). Thermal noise dominated astrometric errors are calculated to be $<$0.07~mas, depending on maser brightness. Proper motion errors are derived from the linear least squares fitting with values of 0.01--0.20~mas~yr$^{-1}$, which are dependent on the number of epochs in the fitting and the astrometric errors of the maser features. Typical rms noise in the line-free channels of the image cubes is $\sim$5~mJy~beam$^{-1}$. Inclination angles $i$ are relative to the plane of the sky. Brightness $I$ and line-of-sight (LOS) V$_{\text{LSR}}$ values are the averages of all detections. Only maser features detected at least in three consecutive epochs are listed, with IDs $\leqslant$92 referring to features detected in all five. Features used for fitting the arcs are indicated in the last column. (This table is also available in plain text ASCII format on the journal website.)
\end{table*}
\end{document}